\documentclass[preprint,amsmath,amssymb,aps]{revtex4-1}
\usepackage{graphicx}
\usepackage{dcolumn}
\usepackage{bm}
\usepackage{float}
\usepackage{color}
\usepackage{amssymb}
\usepackage{tikz}
\usetikzlibrary{shapes}
\usepackage{xcolor}
\usepackage{subfig}
\usepackage[percent]{overpic} 

\begin{document}

\title{Insights into bootstrap percolation: its equivalence with k-core
  percolation and the giant component}

\author{Mat\'ias A. {Di Muro}}
\affiliation{Instituto de Investigaciones F\'isicas de Mar del Plata
  (IFIMAR)-Departamento de F\'isica, Facultad de Ciencias Exactas y
  Naturales, Universidad Nacional de Mar del Plata-CONICET, Funes
  3350, (7600) Mar del Plata, Argentina.}
\email{mdimuro@mdp.edu.ar}

\author{Lucas D. Valdez}
\affiliation{Center for Polymer Studies, Boston University, Boston,
  Massachusetts 02215, USA} 

\author{Sergey V. Buldyrev}
\affiliation{Department of Physics, Yeshiva University, 500 West 185th
  Street, New York, New York 10033, USA}
\affiliation{Politecnico di Milano, Department of Management, Economics and Industrial Engineering,
Via Lambruschini 4, BLD 26, 20156 Milano, Italy}

\author{H. Eugene Stanley}
\affiliation{Center for Polymer Studies, Boston University, Boston,
  Massachusetts 02215, USA}
  
\author{Lidia A. Braunstein}
\affiliation{Instituto de Investigaciones F\'isicas de Mar del Plata
    (IFIMAR)-Departamento de F\'isica, Facultad de Ciencias Exactas y
    Naturales, Universidad Nacional de Mar del Plata-CONICET, Funes
    3350, (7600) Mar del Plata, Argentina.}
  \affiliation{Center for Polymer Studies, Boston University, Boston,
    Massachusetts 02215, USA}

\begin{abstract} 
    
\noindent 
K-core and bootstrap percolation are widely studied models that have
been used to represent and understand diverse deactivation and
activation processes in natural and social systems. Since these models
are considerably similar, it has been suggested in recent years that
they could be complementary. In this manuscript we provide a rigorous
analysis that shows that for any degree and threshold distributions
heterogeneous bootstrap percolation can be mapped into heterogeneous
k-core percolation and vice versa, if the functionality thresholds in
both processes satisfy a complementary relation. Another interesting
problem in bootstrap and k-core percolation is the fraction of nodes
belonging to their giant connected components $P_{\infty b}$ and
$P_{\infty c}$, respectively. We solve this problem analytically for
arbitrary randomly connected graphs and arbitrary threshold
distributions, and we show that $P_{\infty b}$ and $P_{\infty c}$ are not
complementary. Our theoretical results coincide with computer
simulations in the limit of very large graphs.  In bootstrap
percolation, we show that when using the branching theory to compute
the size of the giant component, we must consider two different types
of links, which are related to distinct spanning branches of active
nodes.

\end{abstract}
  
\maketitle

\section{Introduction}

\noindent Threshold models have been used to theoretically describe
processes of contagion in social, financial, and infrastructure
networks
\cite{granovetter1978threshold,morris2000contagion,guilbeault2018complex,kempe2003maximizing}.
Unlike the classic or simple epidemic models used to describe the
spread of infectious diseases, threshold models require a node to have
multiple transmissions from neighbors before changing from an
inactive-susceptible-dysfunctional state to an
active-infected-functional state, or vice versa.  These processes
exhibit propagation of states as cascades that lead to a first-order
transition of differing magnitudes
\cite{Watts,lee2014threshold,bizhani2012discontinuous,cellai2013critical}.
We can use these models to describe the spread of innovation,
information, and behavior among nodes because they tend to change
their state or behavior after interacting with not one, but a group of
other nodes
\cite{kempe2003maximizing,wang2016dynamics,monsted2017evidence}.  For
example, Centola showed in an online social network experiment that an
individual tends to adopt a behavior after several neighbors exhibit
the same behavior \cite{centola2010spread}. A threshold model is an
activation process when the number of active nodes increases with time
and a deactivation process when it decreases.

K-core percolation is one of the simplest threshold models used to study
the deactivation process \cite{dorogovtsev2006k}. In k-core percolation
all nodes are initially active. A fraction $1-p$ of nodes then becomes
inactive or dysfunctional. The fraction of active nodes after the initial failure, $p$, is the control parameter of the model. Then a recursive rule is applied: if an
active node $i$ has fewer than $k_{c}^\ast$ active neighbors, it becomes
inactive. If $k_c^\ast$ is the same for all nodes, then the process is called
homogeneous k-core percolation, if not, then it is called heterogeneous
k-core percolation \cite{cellai2011tricritical}.  In the k-core process,
when all nodes have a number of active nodes greater than or equal to
the threshold $k_{c}^\ast$, the process reaches a steady state. At this
stage the order parameter of k-core is the fraction of active nodes or
the fraction of nodes that belong to the largest connected cluster or
giant component (GC).

Dorogovtsev {\it et al}.~\cite{dorogovtsev2006k} demonstrated that in the
homogeneous k-core process the giant component equals the fraction of
active nodes, and that it exhibits a first-order transition when
computed for several values of initial failure $p$. In addition, Baxter
{\it et al}.~\cite{Bax_2011} found that for heterogeneous k-core there are
finite clusters of active nodes at the steady state, indicating that the
fraction of nodes belonging to the GC is lower than the total fraction
of active nodes. They also found that for the same set of parameters the
process can exhibit simultaneously a continuous and a discontinuous
transition not observed in homogeneous k-core.

Bootstrap percolation is a simple threshold model often used to study
activation processes \cite{adler1988diffusion,Bax_2010}. In this model
all nodes are initially inactive, except for a fraction $f$ of nodes
that activate spontaneously. Then each inactive node becomes active if
it has at least $k_{b}^\ast$ active neighbors. Analogous to k-core,
when $k_b^\ast$ is the same for all nodes the process is homogeneous
bootstrap percolation, and when it is not, the process is heterogeneous bootstrap
percolation. This activation process continues recursively until a
steady state is reached. Baxter {\it et al}.  \cite{Bax_2010,Bax_2011} found
that the total fraction of active nodes $S_b$ exhibit a first-order
transition at a critical value $f_{c1}$. Using a generating function
formalism \cite{Cal_01,New_01,New_02,Bra_01} they also proposed
equations for computing the fraction of active nodes belonging to the
GC, $P_{\infty,b}$, as a function of $f$, but they did not compare
their results with those of stochastic simulations. We perform
simulations of the bootstrap percolation process, and find that the
equations in Ref.~\cite{Bax_2010,Bax_2011} underestimate the fraction
of nodes that belong to the giant component. Using the generating
function formalism, we find the correct solution for $P_{\infty b}$
and show that Refs.~\cite{Bax_2010,Bax_2011}
disregard some activation events when the giant component is
computed.

Although there are several variants for activation and deactivation
models, such as the Watts threshold model and generalized epidemic
models
\cite{chung2014generalized,miller2015complex,gleeson2007seed,Watts}, we
here focus only on the ``canonical'' processes of k-core and bootstrap
percolation explained above. For an extensive description of these
models see Refs.~\cite{lorenz2009systemic,pastor2015epidemic}.

Baxter {\it et al}.~\cite{Bax_2011} compared heterogeneous k-core and
bootstrap percolation and found that they have different structures of
active nodes, which suggests that these processes cannot map each
other. Miller \cite{miller2016equivalence} indicates that the two
processes are complementary because the behavior of active nodes in
heterogeneous k-core percolation is the same as that of inactive nodes
in heterogeneous bootstrap percolation.  Miller proposes that when
mapping the two processes the relationship of the node thresholds in
k-core and bootstrap percolation must be $k_b^\ast = k-k_c^\ast+1$,
where $k$ is the node degree or the number of node connections. In
addition, Janson proves the relation between these processes in random
regular graphs \cite{janson2009percolation}. However, this relation
has not been proven mathematically for a complex network of any degree
distribution $P(k)$, and for any distribution of the
activation/deactivation thresholds.

We here use a generating function formalism to examine the bootstrap
process theoretically and compare our results with those from
stochastic simulations. In Sec.~\ref{sec.equiv} we describe the
equivalence between k-core and bootstrap percolation for any degree
distribution. In Sec.~\ref{sec.GCMat} we present equations for
computing the GC for bootstrap percolation that include activation
events not taken into account in Ref.~\cite{Bax_2011}). Our
analytical results fully agree with our stochastic
simulations. Finally, in Sec.~\ref{sec.Concl} we present our
conclusions.

\section{Equivalence between Bootstrap and K-core percolation}\label{sec.equiv}  

\noindent
In a k-core percolation process, $k_{c}^\ast$ is the threshold number
of active or functional neighbors below which an active node becomes
inactive. We assume that this threshold follows a cumulative
probability distribution $r_c(j,k)=P(k_c^\ast \leq j \mid k)$
\cite{di2017cascading}, where $k$ is the degree of the node and
$k_c^\ast$ is its functionality threshold in k-core percolation. The
function $r_c(j,k)$ is the probability that a node with degree $k$ has
a threshold $k_c^\ast$ lower than or equal to $j$. We assume that at
the beginning of the k-core process, the threshold of any node is not
larger than its degree. Thus, initially, the system is stable, which
is different from the original definition of k-core
\cite{dorogovtsev2006k}. We assume that as a result of the initial
attack, a fraction $f=1-p$ of nodes are destroyed and this initiates
the process of cascading failures at the end of which only the
fraction of active nodes $S_c(p)$ is left.

To connect k-core and bootstrap percolation, we first describe how to
use the generating function formalism to calculate the fraction of
active nodes in the k-core process.

When the process is in a steady state, if we follow a randomly chosen
link in one direction we will end up at a node which we will call
``target'', while in the opposite direction we will reach a node which
we will define as ``root''. In k-core percolation we then define $Z_c$
to be the probability of reaching a target node with at least
$k_c^*-1$ outgoing active neighbors when following a link chosen at
random. Here an outgoing neighbor node of the target node is any
neighbor node other than the root node. Note that root is assumed to
be in the active state, otherwise the node with $k_c^\ast-1$ outgoing
active neighbors must fail according to the rules of the k-core
percolation.  Following Ref.~\cite{di2017cascading}, the fraction of
active nodes $S_{c}$ in the steady state for an initial failure of a
fraction of $1-p$ nodes is
\begin{equation}\label{WWc}
S_{c} =p\Psi_c(Z_c,1-Z_c).
\end{equation}
Here $\Psi_c(Z_c,1-Z_c)$ is the probability that a random node has a
number of active neighbors greater than or equal to its threshold. As
a function of two arguments, $x$, and $y$, $\Psi_c(x,y)$ is defined as
\begin{equation}
  \Psi_c(x,y)=\sum_{k=0}^\infty P(k)\sum_{j=0}^k\binom{k}{j}r_c(j,k)x^jy^{k-j}.
  \label{e:Psi_c} 
\end{equation}
The parameter $Z_c$ satisfies a recursion equation, 
\begin{eqnarray}\label{ZZc}
Z_c =p\Phi_c(Z_c,1-Z_c),
\end{eqnarray}
and  
\begin{equation}
\Phi_c(x,y)=\sum_{k=1}^\infty \frac{k P(k)}{\langle
  k\rangle}\sum_{j=0}^{k-1}\binom{k-\
  1}{j}r_c(j+1,k)x^{j}y^{k-j-1},
\label{e:Phi_c}
\end{equation}
where $\langle k\rangle=\sum_{k=1}^\infty kP(k)$ is the average degree of the network.

On the other hand, in the bootstrap percolation process a fraction $f$
of nodes becomes active at the beginning of the process. We call these
nodes that activate spontaneously the ``seed'' nodes, because they trigger
the activation cascade of the inactive nodes, forming extensive
branches of active nodes. Naturally, we define nodes that were not
activated initially as ``non-seed''.

After the initial activation, a non-seed
node with a degree $k$ becomes active if the number of its active
neighbors $j$ satisfies $j \geq k_b^\ast$, where $k_b^\ast$ is the
bootstrap functionality threshold.
Similar to k-core percolation, the cumulative distribution of the
bootstrap activation threshold is $r_b(j,k)=P(k_b^\ast \leq j |k)$.

\begin{figure}[ht]
\vspace{0.2cm}
\begin{center}
  \includegraphics[width=0.9\textwidth]{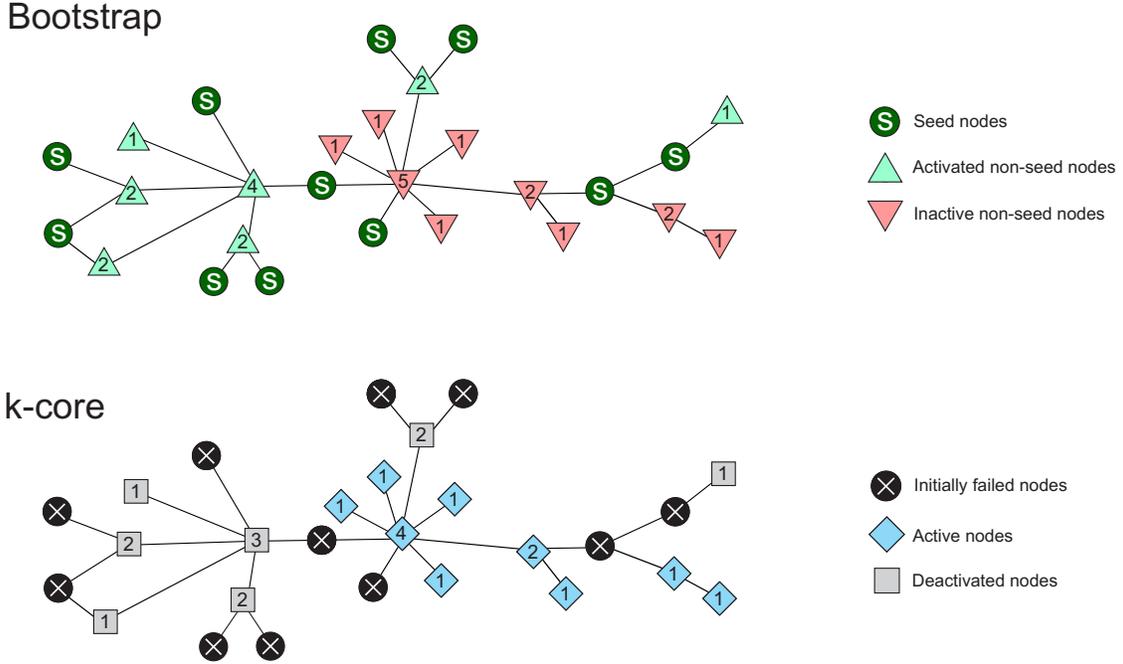}
\end{center}
\vspace{-0.5cm}
\caption{Bootstrap and k-core percolation processes developed on
  a randomly generated network. The bootstrap threshold for a node
  with degree $k$ is $k_b^*=\lfloor{k/2}\rfloor+1$, where
  $\lfloor{x}\rfloor$ denotes the integer part of $x$. If we set the
  seed nodes in bootstrap as initially failed nodes in k-core, then if
  $k_c^*=k-\lfloor{k/2}\rfloor$, according to Eq.~(\ref{thresh}), then
  both process are complementary. All non-seed nodes activated in
  bootstrap coincide with deactivated nodes in k-core, and all
  non-seed nodes that were not activated in bootstrap are the same as
  those that remained active in the k-core process. The numbers
  indicate the thresholds corresponding to each process. Figure
  adapted from Ref.~\cite{miller2016equivalence}}\label{gephi}
\end{figure}

It can be shown that the process of activation of nodes in the
bootstrap percolation with the fraction of seeds $f$ is equivalent to
deactivation of nodes in the complementary k-core percolation process,
in which the initial failure destroyed a fraction $1-p=f$ of nodes, if
their thresholds are complementary. Thus, the seed nodes in bootstrap
play the role of initially failed nodes in k-core. Furthermore, this
implies that the nodes that are active in bootstrap percolation are
inactive in k-core and vice versa.  We will call such nodes b-active
and c-inactive and b-inactive or c-active,
respectively. Fig.~\ref{gephi} illustrates that bootstrap and k-core
are complementary, when both process develop on the same graph with
complementary thresholds.

It can be shown that $Z_c=1-Z_b$, where $Z_b$ is the probability of
reaching, following a random link, a seed node or a non-seed with
$k^*_b$ outgoing links leading to activated nodes. Analogously, it can
be shown that $S_b=1-S_c$, where $S_b$ is the fraction of the active
nodes in the bootstrap percolation. We will provide a rigorous proof
of this equality by deriving the equations of bootstrap percolation
using the k-core equations. For this purpose we have to connect first
the thresholds distributions $r_c(j,k)$ and $r_b(j,k)$.


We have established that the activation of b-inactive nodes is
equivalent to the deactivation of c-active nodes. The condition of
activation is that the number of b-active neighbors, $j_b$ of a
b-inactive node with degree $k$ satisfies $j_b\geq k_b^\ast$. From the
point of view of the k-core percolation, this node has $j_c=k-j_b$
c-active neighbors, and the condition of its deactivation is that $j_c
=k-j_b< k_c^\ast$, or $j_b>k-k_c^\ast$ or $j_b\geq
k+1-k_c^\ast$. Since this last inequality must coincide with
$j_b\geq k_b^\ast$, then,
\begin{equation}\label{thresh}
k_b^\ast=k+1-k_c^\ast.
\end{equation}

This simple equality shows how the thresholds of both process are
related depending on the degree of the nodes. Note that this relation
indicates that in non-regular graphs, the complementary process of
homogeneous bootstrap percolation is heterogeneous k-core
percolation. Likewise, homogeneous k-core percolation is the
complement of heterogeneous bootstrap percolation.

Now we are in conditions to establish a connection between the
threshold distributions of both processes. Using Eq.~\ref{thresh},
\begin{eqnarray}
r_c(j,k)&=&P(k_c^\ast \leq  j  |k),\nonumber\\
r_c(j,k)&=&P(k+1-k_b^\ast \leq  j  |k),\nonumber\\
r_c(j,k)&=&P(k_b^\ast \geq k+1-j  |k),\nonumber\\
1-r_c(j,k)&=&P(k_b^\ast < k+1-j  |k),\nonumber\\
1-r_c(j,k)&=&P(k_b^\ast \leq k-j  |k) \equiv r_b(k-j,k).\nonumber
\end{eqnarray}

Thus, we obtain the relation between the threshold distributions,

\begin{equation}\label{rr}
r_c(j,k)=1-r_b(k-j,k).
\end{equation}

We will show that when the threshold distributions for k-core and
bootstrap percolation satisfy Eq.~(\ref{rr}) then both processes are
complementary.
  
Finally, we will derive the equations for the bootstrap percolation
using Eqs.~(\ref{e:Phi_c}), (\ref{e:Psi_c}) and (\ref{rr}) using the
k-core percolation as a starting point. Note that Eq.~(\ref{ZZc}) can be
rewritten using Eq.~(\ref{rr}) as
\begin{equation}
Z_c=(1-f)\Big(1- \sum_{k=1}^\infty \frac{k P(k)}{\langle
  k\rangle}\sum_{j=0}^{k-1}\binom{k-\ 1}{j}r_b(k-j-1,k)Z_c^{j}(1-Z_c)^{k-j-1}\Big)
\end{equation}  

Introducing new summation index $i=k-(j+1)$ and using the
symmetry of binomial coefficients,
we see, that this equation is equivalent to
\begin{equation}
1-Z_c=f+(1-f)\sum_{k=1}^\infty \frac{k P(k)}{\langle
  k\rangle}\sum_{i=0}^{k-1}\binom{k-\ 1}{i}r_b(i,k)Z_c^{k-i-1}(1-Z_c)^{i}
\end{equation}  

Introducing
\begin{equation}
\Phi_b(x,y)=\sum_{k=1}^\infty \frac{k P(k)}{\langle
  k\rangle}\sum_{i=0}^{k-1}\binom{k-1}{i}r_b(i,k)x^{i}y^{k-i-1},
\label{e:Phi_b}
\end{equation}
we arrive at the following recursive equation
\begin{equation}\label{ZZc2}
  1-Z_c=f+(1-f)\Phi_b(1-Z_c,Z_c).
\end{equation}

If we denote $Z_b=1-Z_c$, then we can write
Eq.~(\ref{ZZc2}) as,

\begin{equation}\label{ZZb3}
  Z_b=f+(1-f)\sum_{k=1}^\infty \frac{k P(k)}{\langle
  k\rangle}\sum_{j=0}^{k-1}\binom{k-1}{j}r_b(j,k)Z_b^{j}(1-Z_b)^{k-j-1},
\end{equation}
or
\begin{equation}\label{ZZb}
  Z_b=f+(1-f)\Phi_b(Z_b,1-Z_b),
\end{equation}
which is the generalization for the analogous equation for the
homogeneous bootstrap percolation~\cite{Bax_2011}. 

Note that the meaning of $Z_b=1-Z_c$ is the probability of a link to
connect {\it any} (active or inactive) root to an already b-activated
node, while $Z_c$ is the probability of a link to connect a c-active
node to a c-active nodes. Indeed, there are three types of links:
those connecting b-active with b-active nodes, those connecting
c-active and b-active nodes and those connecting c-active and c-active
nodes. Since the last category constitute the probability $Z_c$, the
first two together constitute the probability $Z_b=1-Z_c$. Thus, we
conclude that $Z_b$ and $Z_c$ are complementary. The difference in the
meaning of $Z_c$ and $Z_b$ is also reflected in the structure of the
functions $\Phi_b$ and $\Phi_c$ in which the former has a term
$r_b(j,k)$ while the latter has a term $r_c(j+1,k)$.

Using the same techniques as we use for the derivation of $Z_b=1-Z_c$,
it is straightforward to show that $S_b=1-S_c$ and then obtain the
final equation for the active nodes in the bootstrap percolation:
\begin{equation}\label{WWb}
S_{b} =f+(1-f)\Psi_b(Z_b,1-Z_b),
\end{equation}
where
\begin{equation}
\Psi_b(x,y)=\sum_{k=0}^\infty
P(k)\sum_{j=0}^k\binom{k}{j}r_b(j,k)x^jy^{k-j}.\\ 
\end{equation}
This concludes the proof of the complementary of the k-core
percolation and bootstrap percolation with complementary threshold
distributions.

\section{Giant Component equation for bootstrap percolation}\label{sec.GCMat}

\noindent
In this section, we will generalize the equations for the size of the
giant component (GC) in heterogeneous k-core and bootstrap percolation
presented in Ref.~\cite{Bax_2010,Bax_2011} using a different
notation, and then we show that for the bootstrap percolation, their
equations underestimate the size of the GC. The source of this
discrepancy is the difference in the meaning of $Z_b$ and $Z_c$ as we
have shown in the previous section. In the k-core percolation, we
denote as $\alpha_c$ the probability that a randomly selected link,
originating at an active node, leads to the GC of active nodes, and
obviously this probability is less than or equal to $Z_c$. The
probability that a link coming from an active root node lead to an active target node
but not to the GC is $Z_c-\alpha_c$. Thus, the probability that a target node
with a degree $k$ and $j$ outgoing links leading to active neighbors,
is not connected to the GC is
\begin{equation}
\binom{k-1}{j}(Z_c-\alpha_c)^j(1-Z_c)^{k-j-1}.
\end{equation}  
Summing up all these terms for different $k$ and $k_c^\ast$ and after
taking into account the probability of reaching a node with degree $k$
through a random link and the distribution of the thresholds $r_c$, we
conclude that the total probability that any node to which we arrive
by a random link is active but not connected to the GC is
\begin{equation}
  Z_c-\alpha_c=p\Phi_c(Z_c-\alpha_c,1-Z_c).
  \label{e:alpha_c}
\end{equation}  
This equation can be solved together with Eq.~(\ref{ZZc}) for any initial survival probability $p$.
Note that if $r_c(1,k)=0$ for any $k>0$, this system of equations always have
a solution $\alpha=Z_c$ because if $r_c(1,k)=0$, each term of Eq.~(\ref{e:alpha_c}) has a factor $(Z_c-\alpha_c)$, and, thus, the first argument of the function $\Phi_c(Z_c-\alpha_c,1-Z_c)$
can be factored out.
The final equation for the size of the GC, $P_{\infty,c}$, can be written as the probability of randomly choosing an active node, $S_c$, minus the probability of choosing at random an active node with no links connected to the GC, which is $p\Psi_c(Z_c-\alpha,1-Z_c)$. Thus
\begin{equation}
  P_{\infty,c}=S_c-p\Psi_c(Z_c-\alpha,1-Z_c).
  \label{e:P_c}
\end{equation}  
Therefore, if there are no nodes with $k_c^\ast=1$ and no autonomous
nodes ($k_c^\ast=0$), we have $P_{\infty,c}=S_c$, which means that all
active nodes are part of the GC.

Now we will turn to the derivation of the equation for the giant
component in the bootstrap percolation. For brevity we will drop
subscript $b$ in all the equations. Naively, one could expect that the
same equations (\ref{e:alpha_c}) and (\ref{e:P_c}) with small
modifications would work for the bootstrap:
\begin{equation}
  Z-\alpha =f G_1(1-\alpha) +(1-f)\Phi(Z-\alpha,1-Z)
  \label{e:alpha-wrong}
\end{equation}
and
\begin{equation}
  P_\infty =S- [f G_0(1-\alpha) +(1-f)\Psi(Z-\alpha,1-Z)],
 \label{e:P-wrong} 
\end{equation}
where $G_0(x)$ and $G_1(x)$ are the standard generating functions of the network
degree distribution \cite{New_02}:
\begin{eqnarray}
  G_0(x)&=&\sum_{k=0}^\infty P(k)x^k\\
  G_1(x)&=&\sum_{k=1}^\infty \frac{kP(k)}{\langle k\rangle}x^{k-1}
\end{eqnarray}  
The meaning of the terms involving $G_1$ and $G_0$ is the special
treatment of the seed nodes, which are active by default. Thus for
them the classical percolation equations are applied.  Our computer
simulations for a random regular network with degree  $k=3$ and
activation thresholds 2 and 3 ($k=3$ and $k^\ast=2$, $k^\ast=3$) do not agree with these
equations (see Fig.~\ref{RRok}).

To understand the origin of this discrepancy, we need to recall the
definition of $Z$ in bootstrap percolation, which is the probability
of connecting a root node with a seed node, or with an already active
non-seed node. By ``already active node'' we mean a node, whose
activation was triggered by its outgoing neighbors, or what is the
same, that at least $k^*$ of its outgoing links lead to active
nodes. Thus, the activation of the target node does not depend on the
root node, unlike in the k-core percolation, where the root must be
always active. Nevertheless, if the target non-seed node has $j=k^\ast
-1$ outgoing links leading to active neighbors and also the root is
active, the target should be also active, and this possibility has not
been considered in \cite{Bax_2010}.

\begin{figure}[H]
\begin{center}
  \includegraphics[width=0.9\textwidth]{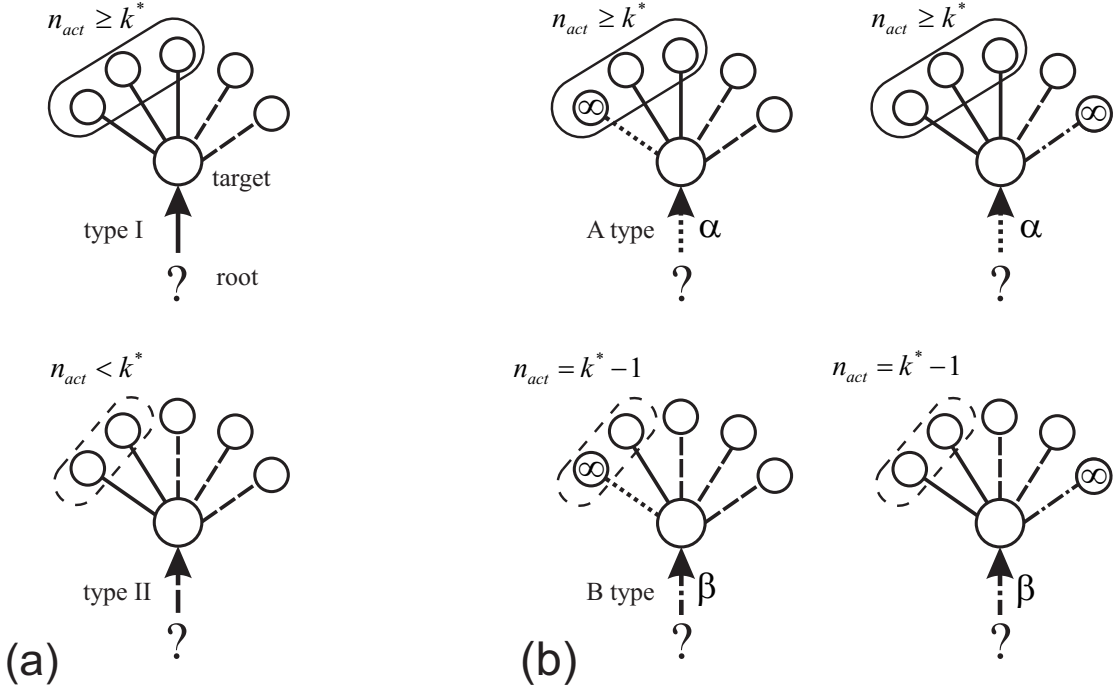}
\end{center}
\vspace{-0.5cm}
\caption{Schematic representation of some configurations when reaching
  a target node following a random edge, in a random 6-regular network
  and for a threshold $k^*=3$. Solid edges lead to an ``already active
  node'', which has $n_{act}=k^*$ active outgoing neighbors (enclosed by a
  solid line). We say these links are of type I. Dashed edges
  connect to a node with less than $n_{act}=k^*$ active outgoing
  neighbors (enclosed by a dashed line), and we call them links of type II. On the other hand,
  dotted lines are edges of type I that lead to the GC, which we call
  type A links, while dash-dotted lines are type II edges that lead to
  the GC, called type B links. Note that a type B link is connected to
  an active node if two conditions are fulfilled: it must lead to a
  target node with exactly $n_{act}=k^*-1$ active outgoing neighbors
  (if $n_{act}>k^*-1$ it would be a type A link), and also the root node must be
  active. In panel (a) we differentiate two types of links: I (top)
  and II (bottom). In panel (b) we show different configurations that
  lead to the GC for each type of link. On top we show a link of type
  A that leads with probability $\alpha$ to a target node connected to
  the GC, because one of its outgoing links is of type A (top-left) or
  of type B (top-right). These configurations are considered in
  Eq.~(\ref{e:alpha}).  On the bottom, we show a link of type B that
  leads with probability $\beta$ to a target node connected to the GC,
  since one of its outgoing links is of type A (bottom-left) or of
  type B (bottom-right). These configurations are considered in
  Eq.~(\ref{e:beta}).} \label{schm}
\end{figure}

Thus the outgoing links leading to the giant component can be of two
types, A or B. The A type include links leading to seed nodes or to
non-seed nodes with $j\geq k^*$ active outgoing neighbors. The B type
include links leading to nodes with $j=k^\ast-1$ outgoing active
neighbors, which can be activated only if the root is active.  Note
that A is the subset of a broader type of links that we will call
links of type I, connecting any root to b-active nodes which have
probability $Z$. In contrast, links of type B are the subset of nodes
of type II which is the complement of type I and, hence has the
probability $1-Z$ (see Fig.~\ref{schm}). We denote the probabilities
that a link of the A and B type lead to the GC by $\alpha$ and $\beta$
respectively.  Thus $\alpha \leq Z$ while $\beta \leq 1-Z$. The total
probability that a randomly selected link leads to the GC is
$X=\alpha+\beta$. Fig.~\ref{schm} illustrates different cases in which
the chosen edge is linked to the GC, with probability $\alpha$ or
$\beta$ for a random regular network with degree $k=6$ and threshold
$k^*=3$.
Since it is sufficient that at least one of the outgoing links of the
target node leads to the giant component, it will be more convenient
to handle the probability that none of the outgoings links lead to the
GC. Thus, we will use the probability that the link of type I does not
lead to the GC, which clearly is $Z-\alpha$, and also the probability
that the link of type II does not lead to the
giant component, which is $1-Z-\beta$. Note that since we have two
types of links, we will need to solve a system of two recursive
equations to compute the GC.

The recursive equation for $\alpha$ can be obtained
by looking at the status of the target. Indeed, if a target is a seed
with probability $f$, the probability that it is not connected to the
GC by the outgoing links is the same as in classical percolation
theory $fG_1(1-X)$. If the target is not a seed node with probability
$1-f$, the probability, that it is not connected to $GC$ is
$(1-f)\Phi(Z-\alpha,1-Z-\beta)$, which differs from the analogous term
for the k-core percolation in Eq.~(\ref{e:alpha_c}) by the replacement
of $1-Z$ by $1-Z-\beta$. This is because the links leading to nodes with
threshold less than the activation threshold, whose probability is
$1-Z$, can still lead to the GC with probability $\beta$. Thus, $\alpha$
satisfies the recursive relation,
\begin{equation}
  Z-\alpha=fG_1(1-X)+(1-f)\Phi(Z-\alpha,1-Z-\beta).  
\label{e:alpha}
\end{equation}  
Note that this equation coincides with the old Eq. (\ref{e:alpha-wrong}) with a correction term $\beta$, replacing
$1-\alpha$ by $1-X$ and $1-Z$ by $1-Z-\beta$.

The recursive relation for $\beta$ can be obtained by the
following arguments.  Suppose that the random link leads to a target
node with threshold $j=k^\ast-1$, which could be activated by the
root. The probability of this event for given degree $k$ and threshold
$k^\ast$ is the $j$th term of Eq.~(\ref{ZZb3}), with $x=Z$, $y=1-Z$
and $j=k^\ast-1$. The probability that this node is not connected to
the GC is given by the same term with $x=Z-\alpha$ and
$y=1-Z-\beta$. Thus, we can compute the probability that such a node
is connected to the GC as the difference between these two
probabilities. After summing up all the contributions from different $k$
and $k^\ast$ and taking into account the degree and threshold
distributions we have
\begin{equation}
\beta=(1-f)[\Delta\Phi(Z,1-Z)-\Delta\Phi(Z-\alpha,1-Z-\beta)],
\label{e:beta}  
\end{equation}  
where
\begin{equation}
\Delta\Phi(x,y)=\sum_{k=1}^\infty \frac{k P(k)}{\langle
  k\rangle}\sum_{j=0}^{k-1}\binom{k-1}{j}[r_b(j+1,k)-r_b(j,k)]x^{j}y^{k-j-1}.
\label{e:beta2}  
\end{equation}  
Here $r_b(j+1,k)-r_b(j,k)$ is the probability that $k^\ast=j+1$, which
is equivalent to $j=k^\ast -1$. Thus, each term corresponds to a node
that is just one active outgoing neighbor short of being activated,
which will be active if the root is active.
To derive the final equation for the fraction of nodes in the GC,
we use a slightly modified Eq. (\ref{e:P-wrong}) with the correction term
$\beta$:
\begin{equation}
  P_\infty=S-[fG_0(1-X)+(1-f)\Psi(Z-\alpha,1-Z-\beta)].
  \label{e:P}
\end{equation}
 Recall that $X=\alpha+\beta$ is the probability of choosing a random
  link that leads to the GC.

As a simple example we
will illustrate these equations for a random regular network with $k=3$ and
$k^\ast=2$. For this network $P(3)=1$, $r_b(0,3)=r_b(1,3)=0$ and
$r_b(2,3)=r_b(3,3)=1$. Accordingly, we have a system of two algebraic
equations:
\begin{eqnarray}\label{e:ab}
  \alpha&=&Z-f(1-X)^2 -(1-f)(Z-\alpha)^2 \\
  \beta&=&(1-f)2[Z(1-Z)-(Z-\alpha)(1-Z-\beta)],\nonumber
 \end{eqnarray}
which can be solved for any fraction of seed nodes $f$ using Eq.~(\ref{ZZb})
\begin{equation}
 Z=f+(1-f)Z^2, 
\end{equation}  
from where we easily obtain $Z=f/(1-f)$.

For a random regular network with $k=3$, Eq.~(\ref{e:P}) is reduced
to
\begin{eqnarray}
 P_\infty=&f&+(1-f)[Z^3+3Z^2(1-Z)+3Z(1-Z)^2]\\
 &-&f(1-X)^3 -(1-f)[(Z-\alpha)^3+3(1-Z-\beta)(Z-\alpha)^2].\nonumber
  \end{eqnarray}
  We verified Eq.~(\ref{e:P}) by simulations for the case of a complex $r_b(j,k)$
  and for a random regular network.

  Note that for the percolation critical point, $\alpha=\beta=0$,
  the Eqs. (\ref{e:ab}) turn into identities with the equation for $\alpha$
  turning into Eq.~(\ref{ZZb}). To find the critical point we can
  present the system (\ref{e:ab}) in a symbolic recursive form
  \begin{equation}
    {\bf x}={\bf A}({\bf x}),
  \end{equation}  
  where ${\bf x}\in R^2$ $(x_1=\alpha,x_2=\beta)$ and ${\bf A}({\bf
    x})$ is a nonlinear operator representing the right-hand side of
  the system of Eqs. (\ref{e:ab}). A sufficient condition of the attractive
  point ${\bf x}=0$ is $|{\bf A}({\bf x})|<|{\bf x}|(1-\epsilon)$,
  where $\epsilon$ is a small positive constant, which in the vicinity
  of zero is equivalent to the condition that the matrix of partial
  derivatives $\partial A_i/\partial x_j|_{{\bf x}=0}$ has absolute values of all its eigenvalues smaller than 1. The critical
  point should be right at the border of the converging and diverging
  behavior, so it should satisfy the condition $\lambda_{max}(f,Z)=\pm
  1$. Together with Eq.~(\ref{ZZb}), it gives the value of the
  critical parameter $f_{c2}$. For example, for a random regular network
  with $k=3$ and $k^*=2$, this condition together with $Z=f/(1-f)$ is
  equivalent to
  \begin{equation} 
  \det\left[ \begin{matrix} 4f-1 &2f \\ 2(1-2f) & 2f-1 \end{matrix}\right] =0,
\end{equation}
which gives $f=1/2$ and $f=1/8$. For $f=1/2$, the second eigenvalue is
2, so the iterations do not converge, while for $f=1/8$ the second
eigenvalue is $-1/4$, so the iterations do converge. Thus, the
critical value for the continuous transition denoted as $f_{c2}$ is
$f_{c2}=1/8$. On the other hand, the old equation
(\ref{e:alpha-wrong}), which neglects $\beta$, gives $f_{c2}=1/4$
(Fig.~\ref{RRok}), predicting a GC much more fragile. A similar
treatment can be applied for $k^*=3$, which gives $f_{c2}\approx 0.344$.

\section{Stochastic Simulations}
We perform stochastic simulations using networks with $N=10^6$ nodes,
to assure a small statistical noise and a negligible probability of loops, so that finite networks can be well approximated by the theoretical results obtained in the limit of infinitely large networks. The networks were generated as randomly connected graphs with a given degree distribution by the Molloy-Reed algorithm \cite{Mol_02}. 

\begin{figure}[ht]
\begin{center}
  \includegraphics[width=0.6\textwidth]{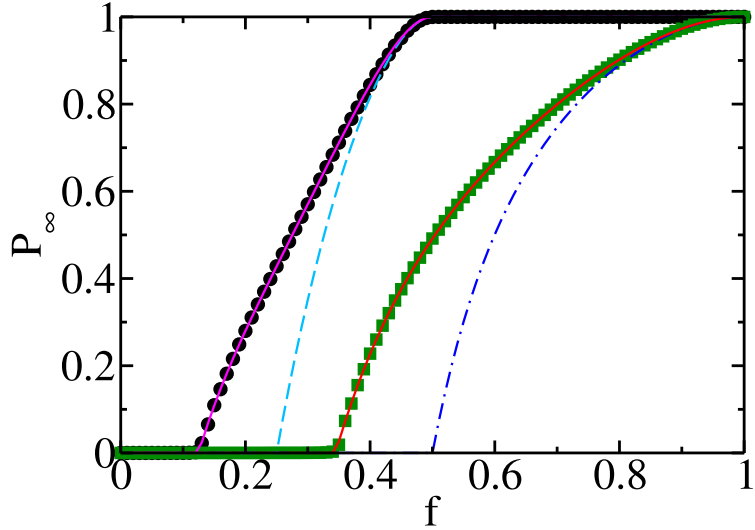}
\end{center}
\vspace{-0.5cm}
\caption{$P_{\infty,b}$ as a function of the fraction of seeds $f$ for
  a random regular network with $k=3$, and thresholds $k^*=2$ (circles, dashed line) and
  $k^*=3$ (squares, dash-dotted line). The symbols represent the
  stochastic simulations with $N=10^{6}$ and the solid lines are the
  prediction of our theory [see Eq.~(\ref{e:P})]. The discontinuous lines represent the results from
  Eq.~(\ref{e:P-wrong}), which underestimate the size of the GC. We can see
  that there is an excellent agreement between the simulations and our equations.}\label{RRok}
\end{figure}

Figure~\ref{RRok} compares the simulation results with 
Eqs.~(\ref{e:P}) and with Eq.~(\ref{e:P-wrong})
corresponding to Ref.~\cite{Bax_2010}. Note that Eq.~(\ref{e:P}) exhibits a good agreement with our stochastic simulations,
while Eq.~(\ref{e:P-wrong}) strongly underestimates the size of GC because it neglects the probability $\beta$.

To test the theoretical equations for the most general case of
heterogeneous k-core and bootstrap percolation, we generate networks
with the Poisson degree distribution with $\langle k \rangle =8$ and a
distribution of the thresholds that satisfy the complementary
condition of Eq.~(\ref{rr}). The form of the cumulative distribution
of bootstrap and k-core percolation are given by:
$r_b(j,k)=F_\gamma(j/k)$, $r_c(j,k)=1-F_{1-\gamma}(1-j/k)$, where
$F_\gamma(x)$ is a fourth-order polynomial of $x$ monotonically
increasing from $F_\gamma(0)=0$, to
\begin{figure}[H]
\vspace{0.2cm}
\begin{center}
   \includegraphics[width=0.48\textwidth]{Fig4a.eps}
  \hspace{0.3cm}
  \includegraphics[width=0.48\textwidth]{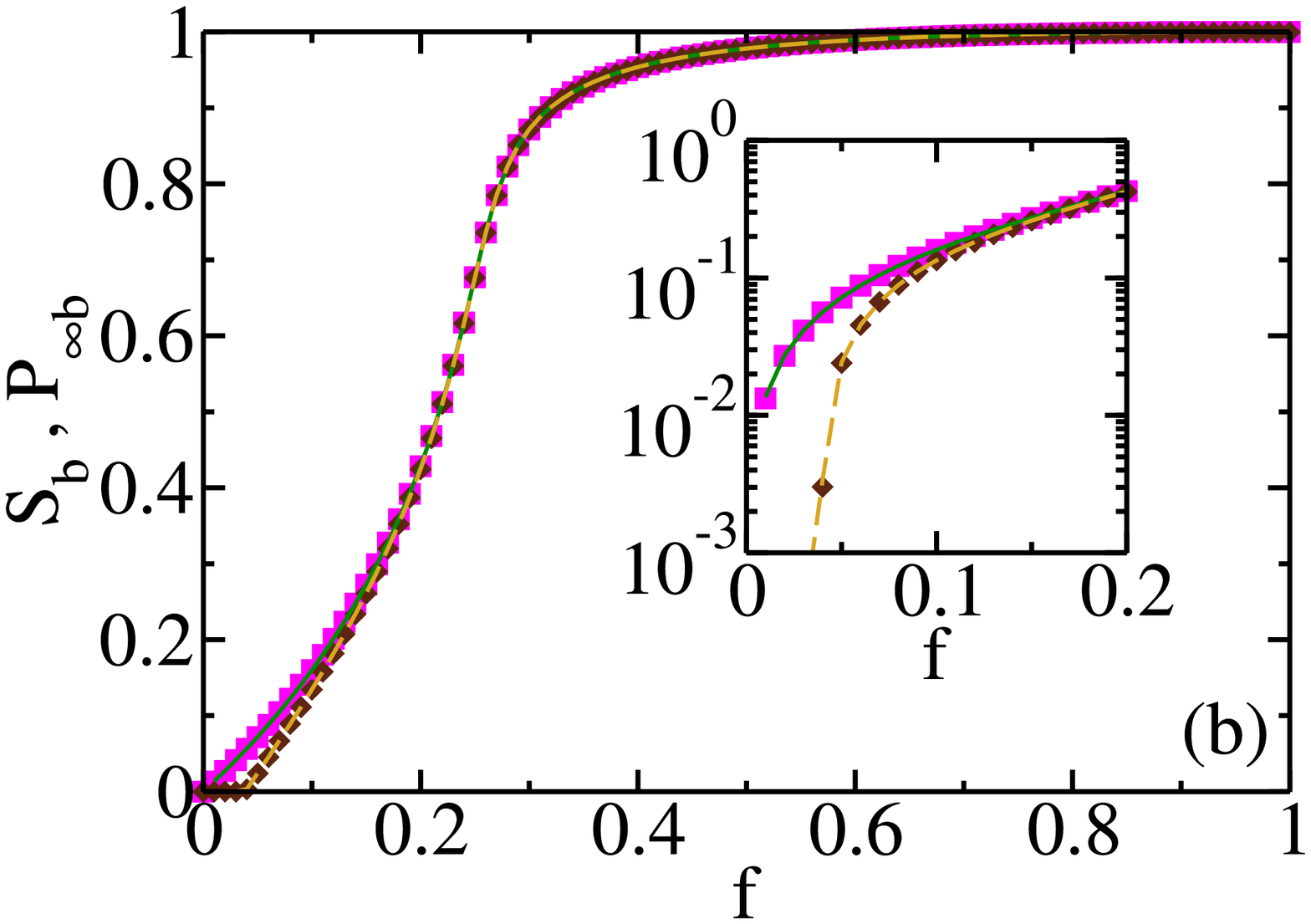}
\end{center}
\begin{center}
  \includegraphics[width=0.48\textwidth]{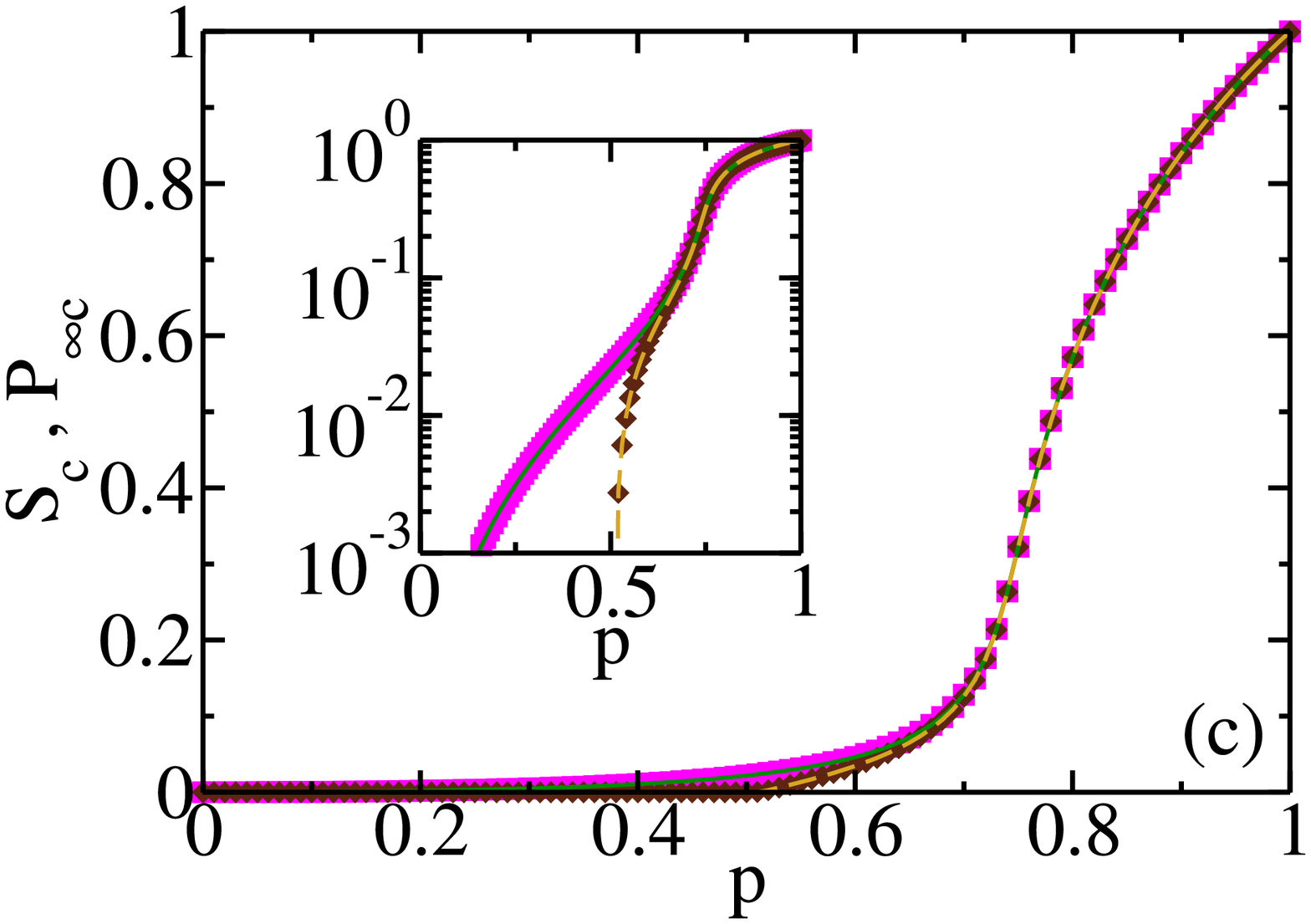}
  \hspace{0.3cm}
  \includegraphics[width=0.48\textwidth]{Fig4d.eps}
\end{center}
\vspace{-0.5cm}
\caption{(a), (c) the fraction of active nodes, $S_c$ (circles,
  squares, solid lines) , and the fraction of nodes in the GC,
  $P_{\infty,c}$ (triangles, diamonds, dashed lines), for the
  heterogeneous k-core percolation as a function of the fraction of
  nodes that survived the initial failure $p$. (b), (d) the fraction
  of active nodes, $S_b$( circles, squares, solid lines), and the
  fraction of nodes in the GC, $P_{\infty,b}$ (triangles, diamonds,
  dashed lines), for the heterogeneous bootstrap percolation as
  function of the fraction of the seed nodes $f$. In the insets we
  plot the same in log-linear scale. The symbols represent stochastic
  simulations while the lines are the results from Eqs.~(\ref{WWb})
  and (\ref{e:P}). All simulations were performed for Erd\H{o}s
  R\'{e}nyi graphs with $\langle k\rangle=8$ and threshold
  distribution functions $r_c(j,k)=F_\gamma(j/k)$,
  $r_b(j,k)=1-F_{1-\gamma}(1-j/k)$, which are polynomials of the fourth power
  with an inflection point at $x=\gamma$: (a) $\gamma=0.5$, (b)
  $\gamma=0.6$, (c) $\gamma=0.4$, (d) $\gamma=0.5$. Note that
  $F_\gamma(x)=1-F_{1-\gamma}(1-x)$, satisfying Eq.~(\ref{rr}) for
  complementary of the k-core and bootstrap thresholds. Accordingly,
  the graphs on the diagonal pairs of panels (a)-(d) and (b)-(c) are
  complementary, that is, $S_c(p)=1-S_b(1-p)=1-S_b(f)$.
}\label{Fig3}
\end{figure}
\noindent $F_\gamma(1)=1$, has a minimum at $x=0$, a maximum at $x=1$
and an inflection point at $x=\gamma$ if $1/3\leq \gamma \leq 2/3$:
\begin{equation}
  F_\gamma(x)=\frac{x^2(18\gamma^2-12\gamma)+x^3(4-12\gamma^2)+x^4(6\gamma-3)}
  {6\gamma^2-6\gamma+1}.
\end{equation}
Note that $F_\gamma(x)$ is invariant under the transformation
$1-F_{1-\gamma}(1-x)$. For nodes with $k=0$, we assume $F_\gamma(0)=1$,
which means that isolated (autonomous) nodes are always
active. Figure~\ref{Fig3} shows the results of computer simulations for
the k-core and bootstrap percolation for different values of $\gamma$.
Depending on the position of the inflection point, we can observe the
emergence of the first order phase transition in both k-core and
bootstrap percolation. We also observe a predicted complementary of
the k-core and the bootstrap percolation
$S_c(p)=1-S_b(1-p)=1-S_b(f)$. In all cases we see excellent agreement
with simulations. Despite the fraction of active nodes of both
processes satisfy a complementary relation, it is clear that the
continuous thresholds of the giant components of these processes do
not complement each other. However, the giant component of inactive
nodes in bootstrap percolation corresponds to the giant component of
active nodes in the complementary k-core percolation process and vice versa.

\section{Conclusion}\label{sec.Concl}
 
We have provided theoretical insights into the bootstrap
percolation process. We prove mathematically that the heterogeneous
bootstrap percolation is the complement of the heterogeneous k-core
percolation for complex networks with any degree distribution in the
thermodynamic limit, as long as the thresholds of the nodes in both
processes complement each other. In particular, in nonregular graphs
we can map a homogeneous bootstrap percolation onto a heterogeneous
k-core percolation, and likewise, k-core homogeneous percolation onto
a heterogeneous bootstrap percolation, because the inactive nodes in
k-core/bootstrap behave the same as the active nodes in
bootstrap/k-core. 

We also develop the equations for the size of the giant component (GC)
in the most general cases of heterogeneous k-core and bootstrap
percolation and confirm them by stochastic simulations. Our equations
for heterogeneous k-core percolation coincide with the equations for a
special case of heterogeneous k-core derived in
Ref.~\cite{Bax_2011}. However, our equations representing the size of
the GC in the bootstrap percolation disagree with the equations
presented in Refs.~\cite{Bax_2010,Bax_2011}. The disagreement comes
from the fact that Refs.~\cite{Bax_2010,Bax_2011} disregard some
branches of active nodes when analyzing the GC in bootstrap
percolation with the generating function formalism. More precisely,
when following a random link that connects a root node and a target
node with degree $k$ and activation threshold $k^*$, it is not
strictly necessary that at least $k^*$ of the $k-1$ outgoing neighbors
of the target be active to ensure its activation. Another possibility
is that only $k^*-1$ outgoing neighbors of the target node be active, and
the root node is also active, it will trigger the activation of the
target node.  We show that the probability of activation of the target
by the root must be explicitly taken into account in order to obtain
the correct equation for the GC in bootstrap
percolation. Nevertheless, to calculate the fraction of active nodes
$S_b$, this activation should not be taken into account since the root
and the target could not mutually depend on each other to be
active. Thus, the equations that represent $S_b$ in
Refs.~\cite{Bax_2010,Bax_2011} are correct.

In the k-core theoretical approach, both the root
and the target nodes are assumed to be active, thus, the root always
acts as a stabilizing neighbor and therefore, the equations from
Ref.~\cite{Bax_2011} predict correctly the size of the GC.

We also found that unlike the fraction of active nodes, the fraction
of nodes belonging to the giant component in both processes do not
satisfy a complementary relation, since these processes generate
different topological structures of active nodes. Indeed, active nodes
in the k-core percolation are inactive in the complementary bootstrap
percolation. However, the giant component of inactive nodes in the
bootstrap coincides with the giant component of active nodes in the
complementary k-core and vice versa.

Our results and theoretical equations here presented can be extended
into networks with multiple layers and can be used to describe the
evolution of the GC of active nodes during this dynamical process.

\section*{Acknowledgments}

\noindent
M.A.D. and L.A.B. thank to UNMdP (Grant No. EXA 853/18), and CONICET (Grant No. PIP 00443/2014) for
financial support. The Boston University Center for Polymer Studies is
supported by NSF Grants No. PHY-1505000, No. CMMI-1125290, and No. CHE-1213217, and
by DTRA Grant HDTRA1-14-1-0017. L.A.B.  acknowledges partial support of this research through 
 DTRA Grant No. HDTRA1-14-1-0017. S.V.B. acknowledges financial support of DTRA, Grant No. HDTRA1-14-1-0017 and the partial support of this
research through the Dr. Bernard W. Gamson Computational Science Center at Yeshiva College.

%

\end{document}